\documentclass[pre,reprint,showpacs,floatfix,amsmath,amssymb]{revtex4-1}

\usepackage{amsmath, amssymb, amsfonts, amsthm}
\usepackage{verbatim}
\usepackage{hyperref}
\usepackage{float} 
\usepackage{soul}
\usepackage{xspace}
\usepackage{cancel}
\usepackage[normalem]{ulem}
\usepackage{mathrsfs}
\usepackage[caption=false]{subfig}
\usepackage{graphicx,color}

\newcommand{\figOne}{
\begin{figure*}[t]
    \centering
\includegraphics[height=1.6in]{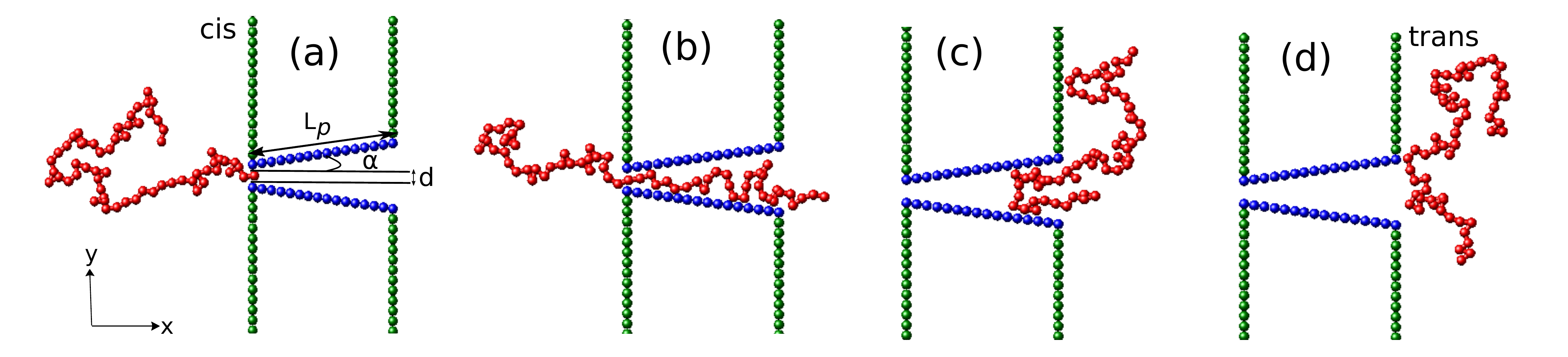}

\caption{Snapshots from the simulation for various stages of the translocation process of a semiflexible polymer with $N = 64$ beads, through a conical channel of length $L_p = 16\sigma$, half apex angle $\alpha$ and channel entrance width $d$. (a) Equilibrium polymer conformation on the {\em cis} side of the channel with one end fixed at the pore entrance. Subsequent polymer conformations (b) midway through the translocation process, (c) showing a possible hairpin formation near the channel exit and (d) when the polymer has successfully translocated to the {\em trans} side of the channel.} \label{fig:1}

\end{figure*}
}

\begin{document}

\title{Driven translocation of a semiflexible polymer through a conical channel in the presence of attractive surface interactions}

\author{Andri Sharma}
\email{ph17007@iisermohali.ac.in}
\author{Rajeev Kapri}
\email{rkapri@iisermohali.ac.in}
\author{Abhishek Chaudhuri}
\email{abhishek@iisermohali.ac.in}

\affiliation{Department of Physical Sciences, Indian Institute of Science Education and Research (IISER) Mohali, Sector 81, S. A. S. Nagar 140306 Punjab, India}

\date{\today}

\begin{abstract}
We study the translocation of a semiflexible polymer through a conical channel with attractive surface interactions and a driving force which varies spatially inside the channel. Using the results of the translocation dynamics of a flexible polymer through an extended channel as control, we first show that the asymmetric shape of the channel gives rise to non-monotonic features in the total translocation time as a function of the apex angle of the channel. The waiting time distributions of individual monomer beads inside the channel show unique features strongly dependent on the driving force and the surface interactions. Polymer stiffness results in longer translocation times for all angles of the channel. Further, non-monotonic features in the translocation time as a function of the channel angle changes substantially as the polymer becomes stiffer, which is reflected in the changing features of the waiting time distributions. We construct a free energy description of the system incorporating entropic and energetic contributions in the low force regime to explain the simulation results. 
\end{abstract}
\maketitle


\section{Introduction}

The translocation of macromolecules through nanopores and channels has been a topic of immense interest since the pioneering work by Kasianowicz et al. in which DNA was successfully translocated through a biological nanopore~\cite{muthukumar2016polymer}. Study of confined polymers and translocation of polymers through geometrical pores paved a way to study sequencing of DNA and RNA, which has a lot of applications in the field of medicine and biotechnology. During the last few decades, a number of experimental and theoretical work has been performed to understand the dynamics of the transportation of such biomolecules~\cite{palyulin2014polymer,branton2010potential,wanunu2012nanopores,howorka2009nanopore,keyser2011controlling,milchev2011single,bezrukov1994counting,kasianowicz1996characterization,sung1996polymer,park1998polymer,muthukumar1999polymer,muthukumar2003polymer,sakaue2007nonequilibrium,sakaue2010sucking,saito2012process,sarabadani2014iso,rowghanian2011force,ikonen2012unifying,ikonen2013influence,ikonen2012influence,bhattacharya2010out,bhattacharya2013translocation,de2010mapping,gauthier2008monte,gauthier2008monte2,huang2014conformations,luo2007influence,luo2008translocation,cohen2011active,cohen2012translocation,sarabadani2018theory,suhonen2018dynamics,katkar2018role,kumar2019complex,suma2017pore,suma2020directional,kumar2019complex,plesa2016direct}. 

The transport of DNA through a nanopore/nanochannel is facilitated by the application of an electric field across the channel, giving rise to an electrostatic force which pulls the DNA into the channel. Due to the reduction in conformational entropy of a polymer in confinement, the geometry of the nanochannel has been found to strongly affect the dynamics of the translocation process. For example, in DNA sequencing using biological nanopores like  $\alpha$-hemolysin and MspA, it has been observed that the shape of channel plays a vital role in determining the ionic current through it \cite{schneider2012dna,sun2014langevin,sun2018simulation,wong2010polymer,meller2001voltage,meller2003dynamics,cohen2012stochastic,kumar2018sequencing}. Compared to the $\alpha$-hemolysin pore, where the cylindrical shape of the beta barrel dilutes the ion current specific to individual nucleotide of the DNA inside the channel, thus making sequencing difficult, the narrow constriction of the cone shaped MspA allows better resolution of current signatures corresponding to individual nucleotide~\cite{derrington2010nanopore,branton2010potential,derrington2010nanopore, venkatesan2011nanopore}. 

\figOne

In a bid to control the transport dynamics of the DNA inside the nanochannel to achieve better sequencing, experimental approaches have shifted to the building of bio-inspired nanopores and nanochannels with similar transport properties~\cite{dekker2010solid,wanunu2007chemically,gershow2007recapturing,luan2012slowing,keyser2006direct}. These artificial solid-state nanochannels not only allow the control of the shape of the channel and tunability of its surface properties, but are also stable with respect to changes in pH, temperature and mechanical oscillations. 
Asymmetrical conically shaped nanopores have been found to have excellent sensing applications, with the nanaopore tip acting as a sensing zone~\cite{mathe2005orientation,harrell2006resistive,lan2011pressure,thacker2012studying,zhou2017enhanced,chen2021dynamics}. Conical nanocapillaries show ionic current rectification at low salt concentrations. Further, chemical modifications of the inner surfaces of a conical nanochannel can lead to reversal in the ion current rectification direction. A combination of a conical nanofunnel with a cylindrical nanochannel, was shown to significantly reduce the threshold external voltage required to trigger the translocation of DNA through the channel~\cite{zhou2017enhanced}. 

Theoretical studies of polymer translocation through conical nanochannels have shown that the escape of a confined flexible polymer from a conical channel is a pore driven process and can proceed without an external force~\cite{nikoofard2013directed,nikoofard2015flexible}. The asymmetric shape of the channel ensures that the polymer has a larger entropic penalty near the constriction which gives rise to an entropic force leading to the escape of the polymer from the larger opening. Further, the passage time is a non-monotonic function of the apex angle of the cone for a given length of the channel. Langevin dynamics studies of flexible polymer translocation through conical channels have shown that translocation is dependent on channel structure and interactions of the polymer with the channel~\cite{sun2018simulation,tu2018conic,sun2021translocation}. Most biopolymers and proteins are however semiflexible with an energy cost associated in bending, characterized by the bending rigidity $\kappa$ of the polymer. The natural question to ask is how the stiffness of the polymer affects the translocation dynamics. It is with this motivation that we characterize the translocation dynamics of a semiflexible polymer through an interactive conical nanochannel.

In this paper, we present a detailed study of translocation dynamics of both flexible and semiflexible polymers through cone shaped nanochannels in the presence of a spatially varying external driving force and surface interactions, using coarse-grained Langevin dynamics simulations. The external force acting along the channel length mimics the voltage driven translocation of polymers, higher at narrower regions of the channel and lower near the larger channel openings. We first report results of the waiting time distributions for a translocating flexible polymer for varying strengths of the external force for a flat channel . The total translocation time $\tau$ decreases with increasing force strength as expected. A break up of the total translocation time into a filling, transfer and escape time provides valuable insight on the translocation dynamics. As the apex angle $\alpha$ (see Fig.~\ref{fig:1}(a)) is varied, the translocation time shows non-monotonic features, consistent with earlier reports~\cite{nikoofard2013directed,nikoofard2015flexible,sun2018simulation,tu2018conic,sun2021translocation}. We comment on the detailed behavior of the waiting time distribution with varying force strengths and $\alpha$. We next present a detailed study of the translocation dynamics of semiflexible polymers with increasing stiffness through this conical nanochannel. The non-monotonic features observed in $\tau$ with varying $\alpha$ differs significantly from that of the flexible polymer. The translocation time is also strongly dependent on the stiffness of the polymer. Further, with increasing forces, the non-monotonic features reduce significantly and total translocation time increases with increasing $\alpha$ for all values of polymer stiffness. We present phase plots of $\tau$ in the $\kappa-\alpha$ plane and mean waiting time distributions to characterize the translocation dynamics in detail. Finally, we provide free energy arguments using a quasi-equilibrium approximation valid at low forces to explain our observations.

The paper is organized as follows. In section 2, we introduce the simulation model and methods where the governing equations are explained. This section also sets the notations used in the paper. In section 3, we present the results and discussion of our simulations for (a) flexible and (b) semiflexible polymer and provide possible explanation for the observed behaviour at low forces via a free energy description. In section 4, we conclude by discussing the importance of our results and possible future directions to extend the domain of physical relevance of this work.

 \begin{centering}
\begin{figure}[h]
\includegraphics[]{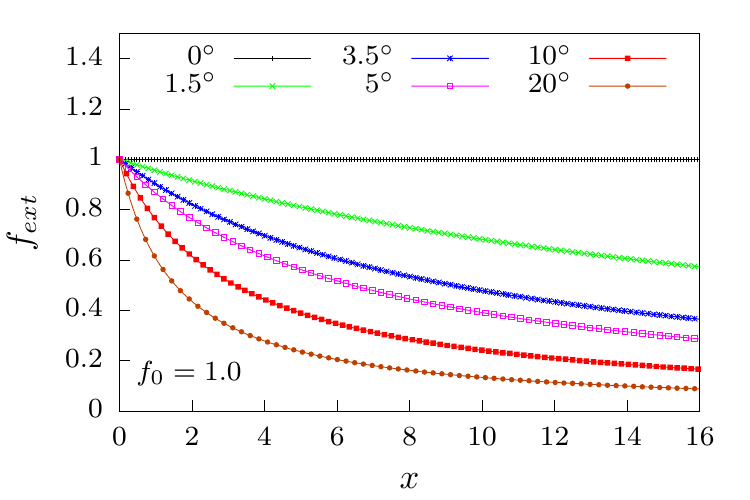} 

\caption{External force $f_{ext}$ as a function of the length of the channel along its axis $x$, plotted for different apex angles $\alpha$ at a fixed $f_0 = 1.0$ according to Eq.~\ref{fy}. With increasing $\alpha$, external force drops rapidly with $x$ as expected.}
\label{f0spatial}
\end{figure}
\end{centering}

\section{Model}

To model the conical channel in two dimensions, we consider the walls of the channel to be made up of two rows of fixed monomer beads of size $\sigma$ and of length $L_p = N^{\prime}\sigma$, where $N^{\prime}$ is the number of monomers making up a wall of the channel. The walls are symmetric about the $x-$axis with the apex angle of the channel $\alpha$ defined with respect to the $x$-axis as shown in the Fig.~\ref{fig:1}. The diameter of the channel at the {\em cis} side is fixed at $d$ to allow only single monomer entry. To separate the {\em cis} and {\em trans} regions of the channel, vertical walls made up of fixed monomers beads are placed along the $y-$ direction. 

The polymer is modeled via a coarse grained bead-spring chain, where non-bonded monomers interact via a short ranged repulsive Lennard-Jones (rLJ) potential given by

\begin{equation}
U_{\textrm{bead}}(r)=
  \begin{cases}
  4 \epsilon \Bigg[ \left(\frac{\sigma}{r}\right)^{12}-\left(\frac{\sigma}{r}\right)^{6} \Bigg]+\epsilon,& r < r_{c}\\ 
  0,& r \geq r_{c} 
  \end{cases}
 \end{equation}
  
\noindent
  where $\epsilon$ gives the strength of the potential. The above truncated and shifted LJ potential has cut-off at  $r_{c}=2^{\frac{1}{6}}\sigma$. The interaction between consecutive monomers of the chain is harmonic with the interaction given by
  \begin{equation}
        U_{\textrm{bond}}=\frac{1}{2}K(r-r_0)^{2},
  \end{equation}
  where $K$ is the spring constant and $r_0$ is the equilibrium separation between consecutive monomers. 
  

\begin{figure*}
    \centering
    \includegraphics[width=\textwidth]{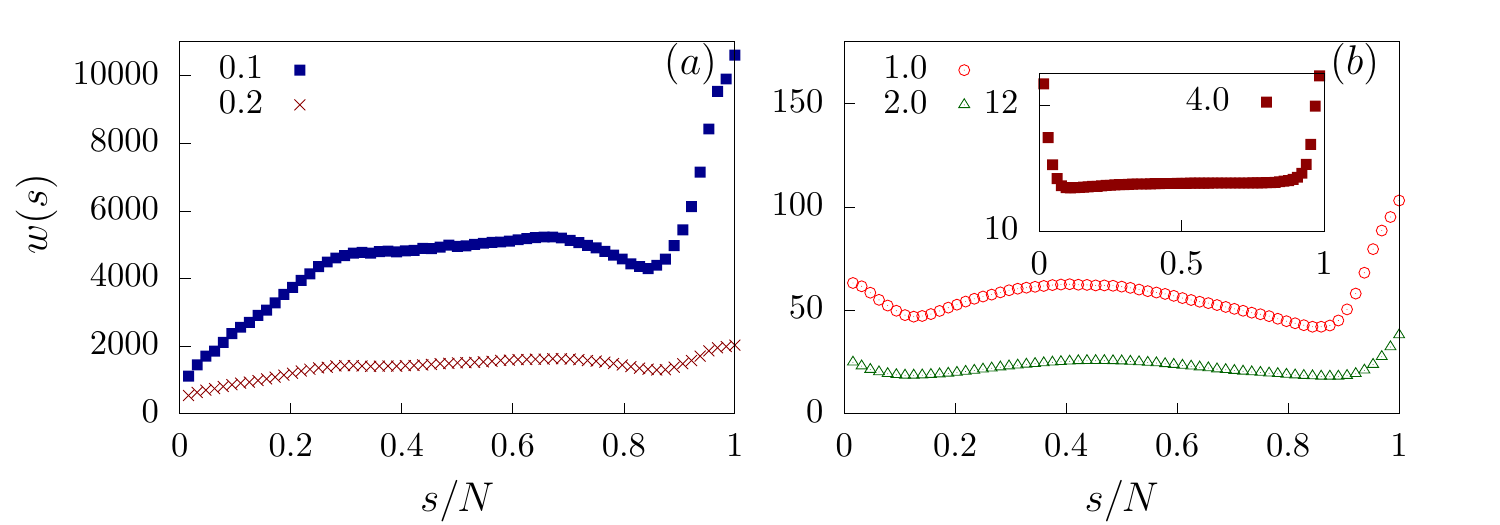}
    \caption{(a) Waiting time distribution $w(s)$ of a flexible polymer through a flat channel ($\alpha = 0^{\circ}$) as a function of scaled translocation coordinate $s/N$, for weak forces $f_0 = 0.1, 0.2$.  (b) Waiting time distribution for strong forces : $f_0 = 1.0, 2.0$. (Inset) $w(s)$ for force values, $f_0 = 4.0$.
}
    \label{fig:revised_newresk0}
\end{figure*}

\begin{figure}
    \centering
    \includegraphics{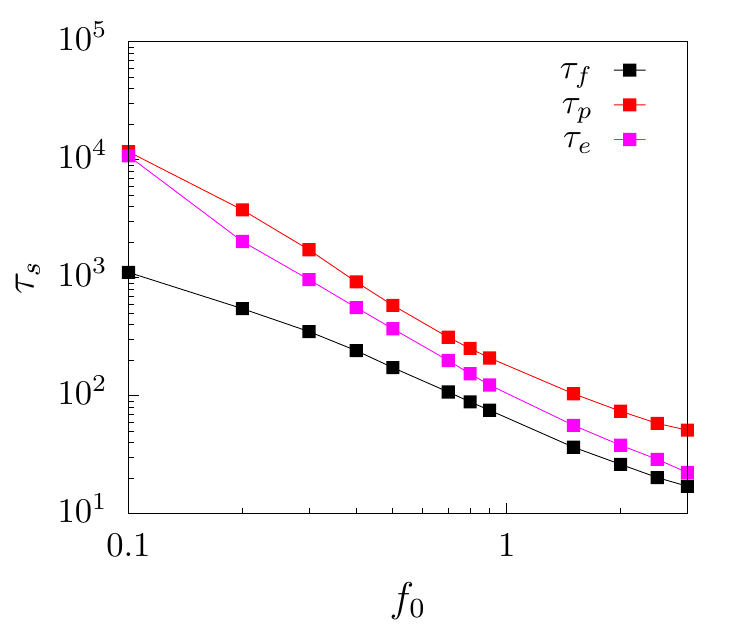}
    \caption{Variation of different time components ($\tau_{s}$) with external force $f_0$ for a flat channel ($\alpha = 0^{\circ}$). Here, $\tau=\tau_{f}+\tau_{p}+\tau_{e}$.}
    \label{tausloglog}
\end{figure}

\begin{figure}
    \centering
    \includegraphics{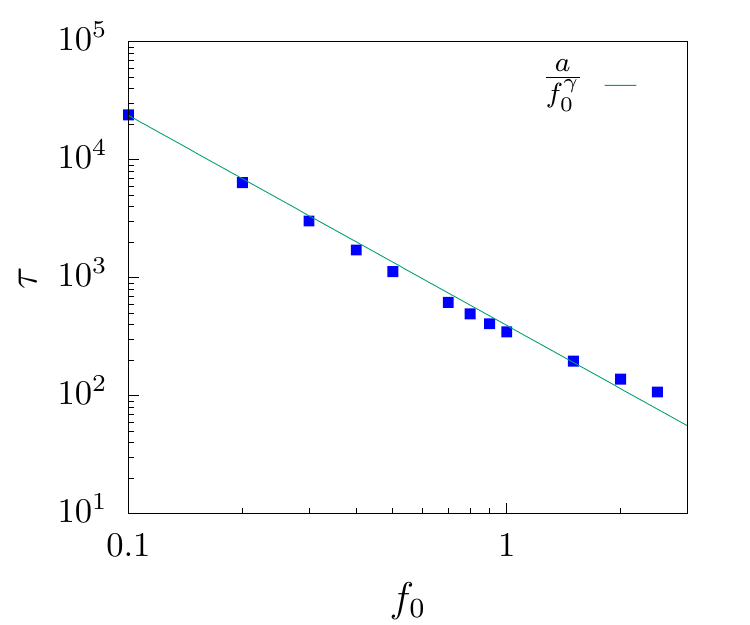}
    \caption{Variation of total translocation time $\tau$ with external force $f_0$ for a flat channel ($\alpha = 0^{\circ}$). $\tau \sim 1/f_0^{\gamma}$ with $\gamma \approx 1.78$.}
    \label{loglog}
\end{figure}

\begin{figure}
\includegraphics[width=0.5\textwidth]{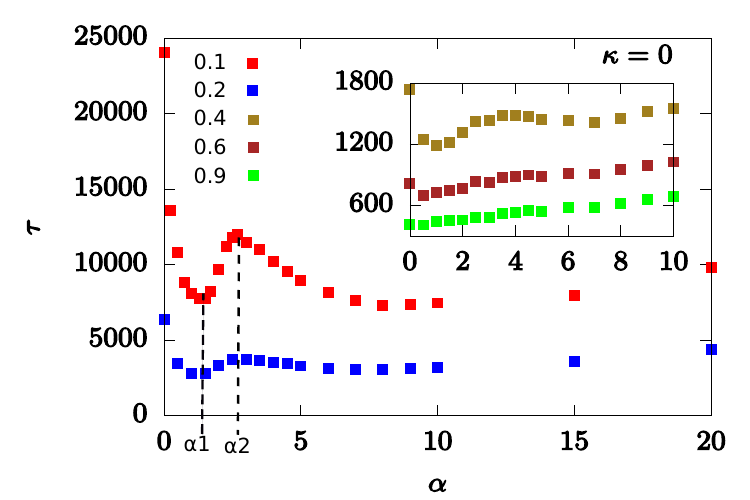}
\caption{Total translocation time ($\tau$) of flexible polymer as a function of apex angle ($\alpha$) for different external forces $f_{0}$. (Inset) Magnified plot of $\tau$ versus $\alpha$ at larger force values showing the disappearance of maxima-minima peaks.}
\label{fig:forwardtransK0_revised}
\end{figure}

\begin{figure}
\includegraphics[width=0.5\textwidth]{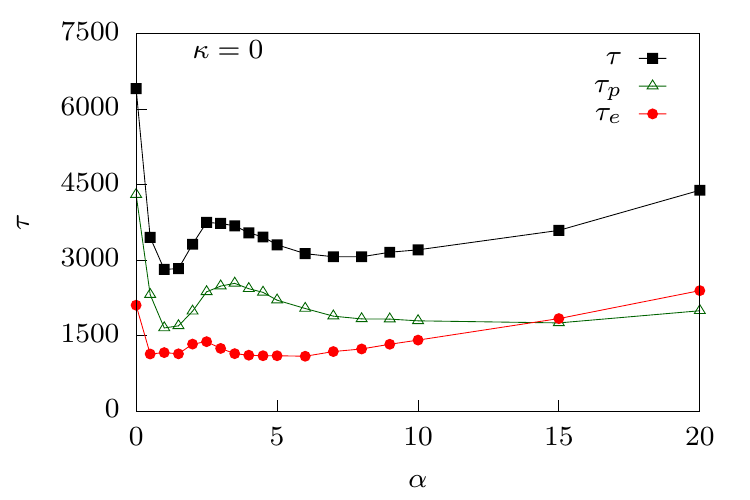}
\caption{Passage time ($\tau_p$) and escape time ($\tau_e$) contributions to the total translocation time for a flexible polymer at a low force value, $f_{0}=0.2$.}
\label{fig:multipletauflexible}
\end{figure}

  To model a semiflexible polymer, an additional bending potential is introduced between consecutive bonds as follows :
  \begin{equation}
    U_{\textrm{bend}}(\theta_{i})= \kappa(1+\cos{\theta_{i}})
\end{equation} 
where $\theta_{i}$ is the angle between the $i^{th}$ and $(i-1)^{th}$ bond vectors. $\kappa$ quantifies the bending rigidity of the polymer. 


The interaction of the polymer beads with the vertical walls are modelled by the same repulsive Lennard-Jones (rLJ) introduced before for the polymer. The surface interaction of the conical channel with the polymer is however attractive, with the attraction between the beads of the polymer and the channel beads given by the standard Lennard-Jones interaction as :
   \begin{equation}
U_{\textrm{channel}}(r)=
  \begin{cases}
  4 \epsilon \Bigg[ \left(\frac{\sigma}{r}\right)^{12}-\left(\frac{\sigma}{r}\right)^{6} \Bigg],& r \leq r_{c}^{lj}. \\
  0,& \text{otherwise} 
  \end{cases}
\end{equation}
where $r_c^{lj} = 2.5\sigma$. 

\begin{figure}
\centering
\includegraphics[width=0.5\textwidth]{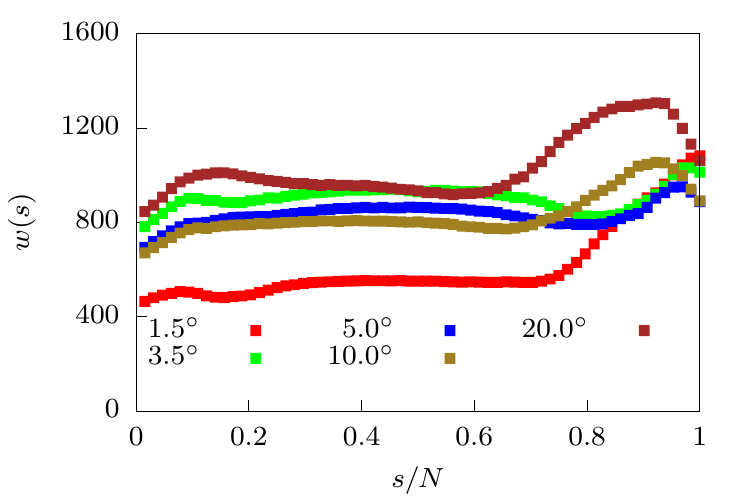}
\caption{ Waiting time distributions ($w(s)$) of a flexible polymer for different $\alpha$ for $f_{0}=0.2$.}\label{fig:forwardresk0}
\end{figure}

In addition to the forces on the polymer due to its interactions with the channel, it experiences an additional external force in the positive $x-$ direction, modeled as a potential gradient, Fig.~\ref{f0spatial} :

\begin{equation}\label{fy}
f_{ext}=\frac{f_{0}d}{(d+2x\tan\alpha)}
\end{equation}
where $f_0$ is the constant force expected in a linear channel with $\alpha = 0^{\circ}$. Evidently, for a conical channel the force decreases along its length.    

The equation of motion for the position of the $i$th monomer of the polymer is given by 
\begin{equation}
m \Ddot{\vec{r_{i}}}=-\zeta\dot{\vec{r_{i}}}-\nabla U_i + \vec{f}_{ext} + \vec{\eta}_i,
\end{equation}
 here m is the mass of the monomer, $U_i$ is the net potential faced by the monomer, $\zeta$ is the friction coefficient and $\vec{\eta}_i$ is a Gaussian random force with $\langle \eta_{i}(t)\eta_{j}(t^{'})\rangle=2k_{B}T\zeta\delta_{ij}\delta(t-t^{'})$ where $k_B$ is the Boltzmann's constant and $T$ is the temperature. The equations are simulated in LAMMPS using the Verlet update scheme. The scales of length, energy and mass are set by $\sigma$, $\epsilon$ and $m$ respectively. This sets the time scale as 
  $\tau_{0}=\sqrt{\frac{m\sigma^{2}}{\epsilon}}$.  In these units, we choose $k_BT = 1.0$, $\zeta = 1.0$, $r_0 = 1.12$, $d = 1.25$ and $K = 10^3 k_B T/\sigma^2$.$f_0$ is varied in the range $0.1-2 \,\,k_BT/\sigma$ and $\kappa$ varies in the range $0-8$. The number of polymer beads $N = 64$ and pore beads $N^{\prime} = 16$ are fixed in our simulations. The time step is set as $\Delta t = 0.001\tau_0$ and all  results presented are averaged over $1500-2000$ independent samples.
  
  We start our simulation by holding the first bead of the polymer at the narrow opening of the channel while the rest of the polymer segment is allowed to relax to its equilibrium conformation outside the channel. The translocation process is tracked by the translocation coordinate $s$. The counter for $s$ starts once a bead of the polymer reaches the {\em trans} side after crossing the channel. The polymer is said to be successfully translocated if $s$ is equal to the total number of  beads $N$ of the polymer and the simulation terminates. The speed with which the polymer traverses the channel gives a detailed idea of what is going on inside the channel. To trace the above effect, we focus on the waiting time $w(s)$, defined as the average time spent by a monomer inside the channel. 

\begin{figure*}
\centering
\includegraphics[width=0.9\textwidth]{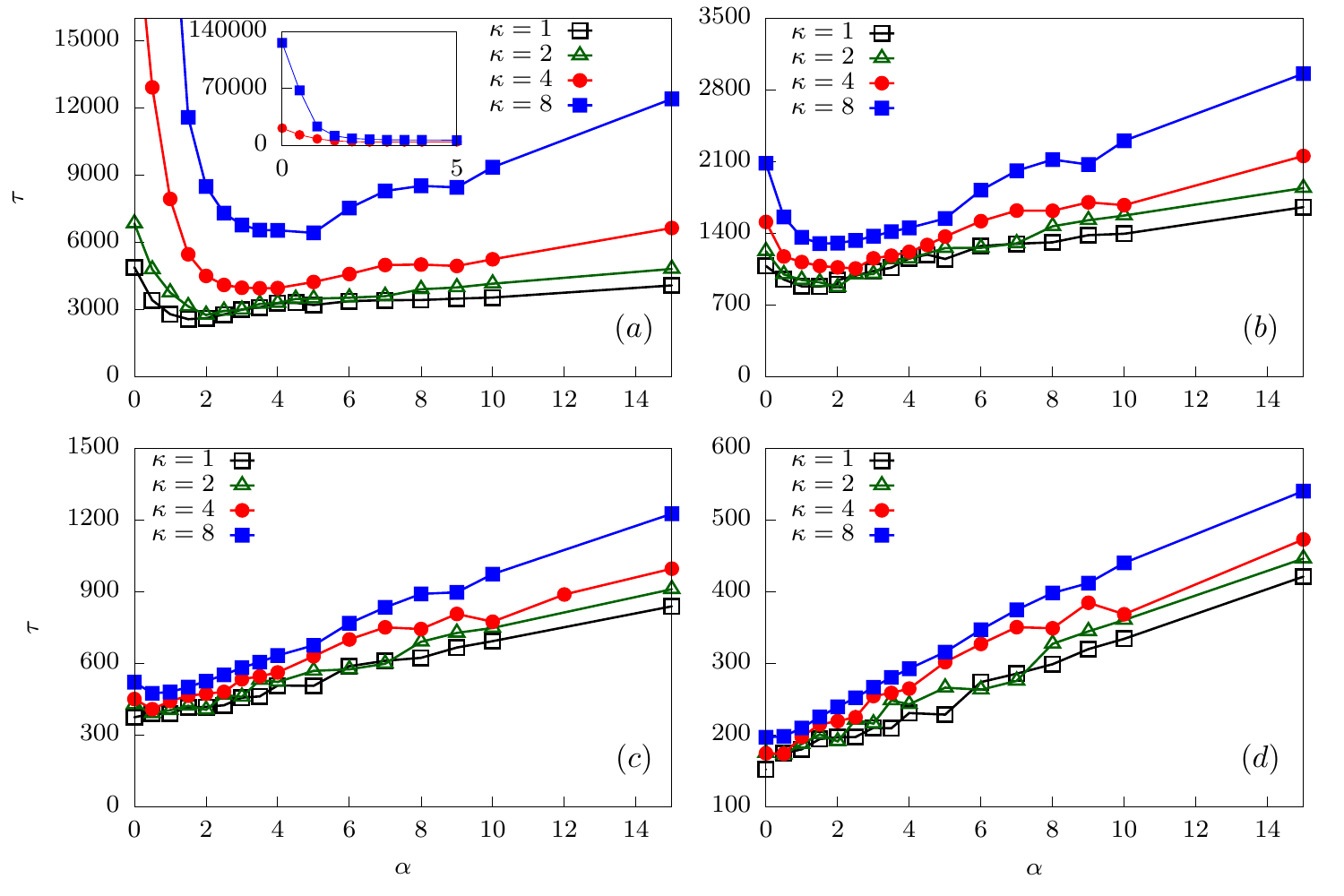}
\caption{Total translocation time versus $\alpha$ for a semiflexible polymer for four different bending rigidties $\kappa = 1,2,4$ and $8$. In (a) the variation is shown for $f_0 = 0.2$. (Inset) Magnified plot at lower $\alpha$ for $\kappa = 1.0$ and $2.0$. As external force is increased : (b) $f_0 = 0.5$ (c) $f_0 = 1.0$ and (d) $f_0 = 2.0$, $\tau$ increases almost linearly with $\alpha$.}
\label{rigidtrans}
\end{figure*}

\section{Results and Discussion}

\subsection{Flexible Polymer}


\begin{figure*}[h]
\centering
\includegraphics[width=\textwidth]{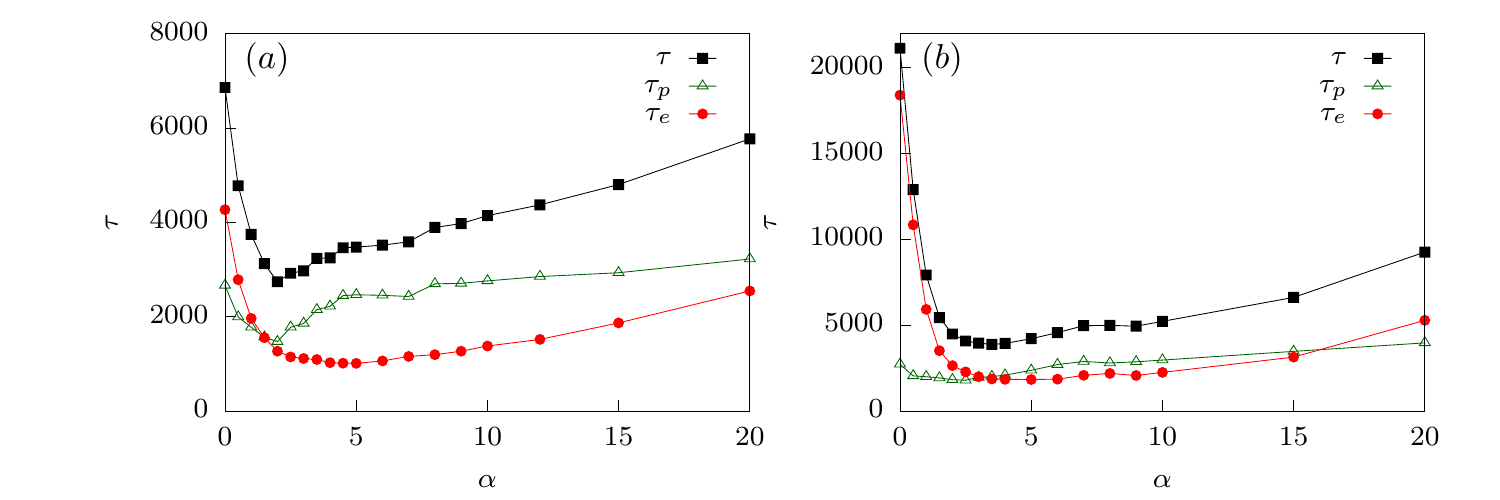}
\caption{Translocation time ($\tau$) as a function of apex angle ($\alpha$) separated into passage time ($\tau_p$) and escape time ($\tau_e$) components for two different values of rigidity of the polymer (a) $\kappa = 2$ and (b) $\kappa = 4$ for $f_0 = 0.2$.}
\label{timecomponent}
\end{figure*}

We first look at the translocation dynamics of a flexible polymer through a flat channel $\alpha = 0^{\circ}$ which serves as a control. As the strength of the external force $f_0$ acting inside the channel is increased ($f_{ext} = f_0$ for a flat channel), the total translocation time ($\tau$) of the polymer decreases. 
To explain this behavior, we look at the waiting time distribution $w(s)$ of the polymer at different values of $f_0$ (Fig.~\ref{fig:revised_newresk0}). Tension propagation theory~\cite{sarabadani2014iso,ikonen2012unifying,ikonen2012influence,bhattacharya2013translocation} accurately predicts the waiting time distribution for a slit (channel of unit size). $w(s)$ is expected to rise with $s$ as the tension due to the external force reaches the end of the polymer in the {\em cis} side. After that the system is in the tail retraction stage and $w(s)$ starts decreasing with $s$. In our system, where the pore is extended and is attractive in nature, we observe several interesting features in $w(s)$ which are different from that of the slit. These arise due to the interplay between surface interactions and the external drive.

For the initial part of the translocation when the polymer enters the pore, the attractive nature of the pore and the external force, sucks the polymer inside the pore. For smaller forces $(f_0 = 0.1, 0.2)$ (Fig.~\ref{fig:revised_newresk0}(a)), $w(s)$ rises with $s$ as the tension slowly reaches towards the end. However, as $f_0$ increases (see Fig.~\ref{fig:revised_newresk0}(b) with $f_0 = 1.0,2.0$), the combined effect of the attractive pore and the external force pulls the initial monomers quicker inside the pore, thus reducing their waiting times. For the lower forces (Fig.~\ref{fig:revised_newresk0}(a)), we observe a flat regime where the $w(s)$ does not change with increasing $s$. Subsequently, once the tension has propagated to the end, the tail retraction part sets in and $w(s)$ starts to fall. Dramatically however, $w(s)$ shows a sharp rise for the end monomers of the polymer. This feature is true for all external force values. As the number of monomers left in the channel becomes lesser than the channel length and keeps decreasing, the net external force on the polymer decreases. Therefore, the attractive nature of the channel becomes a dominant factor for the end monomers resulting in increased waiting times.

The behavior of the waiting times as explained above can be better understood if we divide the total translocation time as $\tau = \tau_f + \tau_p + \tau_e$ where (i) $\tau_f$ is the initial filling time, the time taken by the first monomer of the polymer to reach the exit without returning back into the channel; (ii) $\tau_p$ is the transfer time, the time taken from the exit of the first monomer into the {\em trans} side to the entry of the last monomer from the {\em cis} side; and (iii) $\tau_e$ is the escape
time, the time between the entry of the last monomer in the channel and its escape to the trans-side~\cite{luo2008translocation,kumar2018sequencing}. $\tau_f$ depends strongly on the external drive and the surface interactions which sucks the polymer inside. $\tau_p$ depends on surface interactions and entropy of the polymer segment outside the channel on the {\em trans} side and less on the external force since these monomers see a constant force inside the channel. $\tau_e$ is strongly influenced by the attractive interactions which hold back the polymer as it exits the channel.  In Fig.~\ref{tausloglog}, we have plotted the three translocation times as a function of the force. For low forces, $\tau_f$ is significantly smaller than $\tau_p$ and $\tau_e$, suggesting that translocation is dominated by the transfer and escape dynamics. This explains the initial rise in $w(s)$ for small $s$ as successive monomers spent longer times inside the channel. At larger forces, $\tau_f$ becomes comparable to the transfer and escape times. As explained earlier, the combined effect of surface interactions and external force, pulls the monomer rapidly inside the pore. Therefore, the initial monomers show a dip in the waiting times.

In Fig.~\ref{loglog}, we have plotted the total translocation time of the polymer as a function of the external force. The translocation time falls faster than $1/f_{0}$ for larger forces. This behavior is reminiscent of the effect of attractive pore-polymer interactions which provides an additional pull on the monomer beads as they enter the pore. Such behavior of $\tau$ with $f_0$ for driven polymer translocation through extended channels have been reported earlier~\cite{luo2007influence}.

 We next look at the situation for the conical channel ($\alpha \neq 0$). Since more polymer conformations are possible in the extended part of the channel compared to the constricted region, this shape asymmetry leads to a force of entropic origin which can facilitate movement of the polymer from the {\em cis} to the {\em trans} end. The shape asymmetry therefore influences the translocation process significantly in addition to the effects of the attractive surface interactions of the channel. Further the external force is space dependent (Fig.~\ref{f0spatial}) and decreases as we move from the {\em cis} to the {\em trans} side of the channel. This leads to a global decrease in the total translocation time ($\tau$) when compared to a flat channel. Additionally, $\tau$ displays distinct non-monotonic features as $\alpha$ is varied (Fig.~\ref{fig:forwardtransK0_revised}). We first analyze this non-monotonic behavior for very low forces (say $f_0 = 0.1$).
 
  As $\alpha$ starts to increase, $\tau$ decreases as translocation is facilitated by the additional entropic drive when compared to $\alpha = 0^{\circ}$ which drives the polymer inside the conical pore faster compared to a flat channel. This continues till a certain $\alpha = \alpha_{1}$ (see Fig.~\ref{fig:forwardtransK0_revised}; for $f_0 = 0.1$, $\alpha_{1}\approx 1.5^{\circ}$).  Beyond this apex angle, $\tau$ starts increasing with $\alpha$. Note that with increasing $\alpha$, the attractive interactions along the length of the channel weaken. This lowering of the pull on the polymer near the entrance leads to longer translocation times. It is again useful to split the total translocation time as was done for the flat channel. However, unlike a flat channel, a conical channel has a larger exit which results in folding of the polymer back into the channel (see Fig.~\ref{fig:1}(c)). Therefore, for such channels it becomes convenient to split the translocation time into a passage time and an escape time, $\tau = \tau_p + \tau_e$. Passage time $\tau_p$ for a conical channel is defined as the time taken from the time of entrance of the first bead of the polymer to the time the last bead of the polymer reaches the entrance of the pore from the {\em cis} side. The escape time $\tau_e$ is defined similar to that of the flat channel as the time taken after $\tau_p$ until all the polymer beads have exited the pore from the {\em trans} side. As we see, from Fig.~\ref{fig:multipletauflexible}, the non-monotonicity in $\tau$ for smaller $\alpha$ is completely governed by $\tau_p$. The weakening of the surface interactions near the {\em cis} side is what dominates the behavior.
  
  
  This increase in $\tau$ with $\alpha$ continues till it reaches $\alpha = \alpha_2$. Beyond $\alpha_2$, $\tau$ starts to decrease with $\alpha$. This is a combined effect of the entropic drive and surface interactions. Due to a larger pore exit, a pore which is completely filled with the polymer, experiences a stronger entropic drive. Further, the escape time ($\tau_e$) also decreases as the attractive interactions near the exit becomes negligible and cannot hold back the polymer. With increasing $\alpha$, we see a curious behaviour. At high $\alpha$, $\tau_e$ starts to increase and becomes larger than $\tau_p$ at some $\alpha$. This crossover can be explained by noting that the exit is now large enough to allow folding of the polymer leading to hairpin formations. Escape of the polymer becomes the rate limiting step. Note that beyond a value of $\alpha$, $\tau_p$ essentially saturates as the forces that drive the entry and subsequent passage towards the exit do not change.
  
  At higher forces ($f_{0}>0.4$; see inset of Fig.~\ref{fig:forwardtransK0_revised}), $\tau$ increase with $\alpha$. To understand this behavior, note that for higher forces, the surface interactions do not play a major role. The force along the channel axis is larger for smaller $\alpha$. This ensures that at large force values the translocation process is extremely fast. As $\alpha$ increases, the force along the channel axis decreases (see Fig.~\ref{f0spatial}) and translocation becomes slower.
  
  We further verify our findings for the translocation time by looking at the waiting time distribution of the monomer beads (Fig.~\ref{fig:forwardresk0}). At a force value of $f_0 = 0.2$, we note that as $\alpha$ is increased from $1.5^{\circ}$ to $3.5^{\circ}$, $w(s)$ for the initial beads increases dramatically. This is a result of weakening attractive interactions near the {\em cis} side as the pore broadens. Note that for the end beads, $w(s)$ for $\alpha = 1.5^{\circ}$ rises sharply because of the attractive interactions at the {\em trans} side which prevent exit of the polymer. This effect is reduced sharply for $\alpha = 3.5^{\circ}$. As $\alpha$ is increased further,  $w(s)$ reduces due to weakening attractive interactions near the exit and a greater entropic drive. At even higher $\alpha = 10^{\circ}$, we start seeing the effects of the folding in of the polymer leading to increased $w(s)$ for all monomer beads when compared to $w(s)$ at lower $\alpha$ values. The dip in $w(s)$ for the last few monomers at large values of $\alpha$ is easily explained. With surface interactions lowered significantly and large part of the polymer already outside the pore leading to a higher entropy, the end monomers are sucked out extremely fast.

\begin{figure*}
\centering
\includegraphics[width=\textwidth]{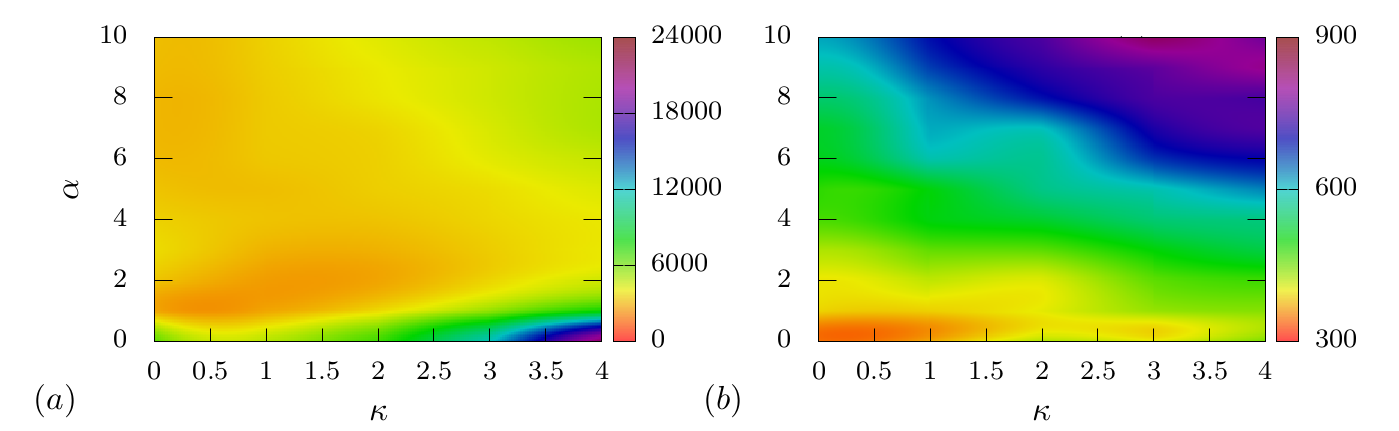}
\caption{Phase plot of total translocation time $\tau$ in the $\kappa - \alpha$ plane for two forces (a) $f_{0}$=0.2. and (b) $f_0 = 1.0$.}
\label{phaseplot_new}
\end{figure*}

\begin{figure*}
\centering
\includegraphics[width=\textwidth]{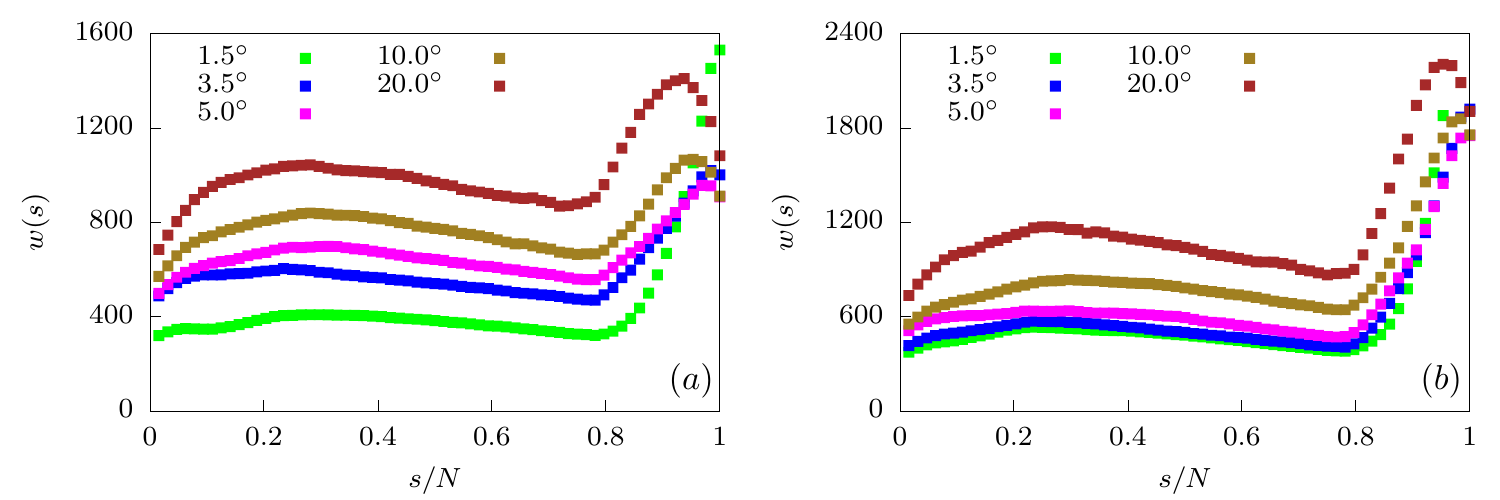}
\caption{Waiting time distributions for various $\alpha$ at (a) $\kappa= 2$, and (b) $\kappa= 4$ for $f_{0}=0.2$. The distributions show distinct differences as the bending rigidity of the polymer is increased.   
}\label{rk1248f02}
\end{figure*}

\subsection{Semiflexible Polymer}
We now discuss the translocation dynamics of a driven semiflexible polymer through the conical channel. We first look at the total translocation time as function of the apex angle for different values of polymer rigidity and external driving force. In Fig.~\ref{rigidtrans}(a), we have plotted $\tau$ as function of $\alpha$ for a low external force value ($f_0 = 0.2$). The variation of the translocation time with apex angle still shows a non-monotonic feature as observed in the case of a flexible polymer. However, unlike that of the flexible polymer, rigid polymers do not show a sharp secondary peak in $\tau$. 
The total translocation time shows an initial decrease with increasing $\alpha$ indicating that as the apex angle increases, it becomes favorable for the polymer to exit the channel. This is a result of the enhanced entropic drive that we discussed for the flexible case. Beyond a critical value of $\alpha$, $\tau$ starts to increase again. Note that the critical $\alpha$ shifts to larger values of apex angle and to higher values of $\tau$ as the rigidity of the polymer is increased (see Fig.~\ref{rigidtrans}(a)). This indicates that translocation becomes more difficult with increasing stiffness of the polymer. This feature is consistent with earlier results of driven translocation of semiflexible polymers through extended channels~\cite{adhikari2013driven,kumar2018sequencing,sarabadani2017driven}. The largely monotonic increase in $\tau$ with increasing $\alpha$ is different from that of the flexible polymer.

The effect of rigidity can be explained by looking at the multiple stages of the translocation process. At lower values of rigidity (say $\kappa = 2$, see Fig.~\ref{timecomponent}(a)), $\tau_p$ plays a more significant role for most $\alpha$ values. This shows that the total translocation time is dictated by the filling of the pore as in the flexible case. However, with increasing rigidity (say $\kappa = 4$, see Fig.~\ref{timecomponent}(b)), $\tau_e$ becomes the rate limiting process. The increased influence of the escape time dynamics happens because the rigid polymer which at this stage fills the channel, keeps deflecting between the two walls of the channel. This is a behavior typical of semiflexible polymers with increased rigidity~\cite{odijk1983statistics}.  The attractive interactions of the two walls keeps the polymer pinned inside the channel for long times, increasing the escape time $\tau_e$. The monotonic increase of $\tau$ with $\alpha$ as opposed to the appearance of a second peak for flexible polymers in the conical channel, can also be explained from the dominance of the escape time dynamics. The competition of the attractive interactions with the entropic drive which results in the lowering of translocation time beyond $\alpha_2$ for flexible polymers is no longer the rate limiting step. Rather, escape time dynamics which are dominated by the attractive interactions of the channel walls with the escaping beads influences the translocation process.

As the external force is increased, the non-monotonicity observed in the variation of $\tau$ with $\alpha$ starts decreasing (Fig.~\ref{rigidtrans}(b-d)). 
At very high forces, $f_{0} \geq 1.0$  (see Fig.~\ref{rigidtrans}(d)), the total translocation time increases monotonically with the apex angle for all rigidities of the polymer. In Fig.~\ref{phaseplot_new} we construct phase plots for $\tau$ in the $\kappa-\alpha$ plane for various values of $f_0$. For small forces $f_0 = 0.2$, $\tau$ is maximum for stiffer chains and smaller apex angles (see Fig.~\ref{phaseplot_new}(a), $\kappa > 3,\,\, 0^{\circ} \leq \alpha \leq 1^{\circ}$). For larger forces $f_0 = 1.0$, the phase plot shows a largely uniform increase in $\tau$ with increasing $\alpha$ for all $\kappa$ values (see Fig.~\ref{phaseplot_new}(b)). The maximum translocation times are observed when both $\kappa$ and $\alpha$ are large.

We look at the waiting time distribution $w(s)$ of the semiflexible polymer as it moves from the {\em cis} to the {\em trans} side of the channel at low forces ($f_0 = 0.2$). In Fig.~\ref{rk1248f02}, we have plotted $w(s)$ for different values of the bending rigidity as the apex angle is varied. For low bending rigidity, $w(s)$ shows an increase with $s$ initially and then a largely flat region at intermediate $s$, for all values of $\alpha$ (Fig.~\ref{rk1248f02}(a)). These results are similar to that for uniformly extended channels. Near the end, the attractive interactions prevent exit of the polymer leading to larger waiting times. For stiffer polymers, the waiting times for all monomers monotonically increase with $\alpha$ which is consistent with the behavior of total translocation time. As the polymer  rigidity increases (Fig.~\ref{rk1248f02}(b)), the entropic gain in moving towards the {\em trans} end of the conical channel becomes lesser. The translocation dynamics of the end monomers is now dominated by the attractive surface interactions. The end segment of the polymer keeps deflecting between the walls and are stuck in either wall for long periods, thereby increasing the waiting times.

\section{Free Energy}

\begin{figure*}
\subfloat[]{\includegraphics[width=0.48\textwidth]{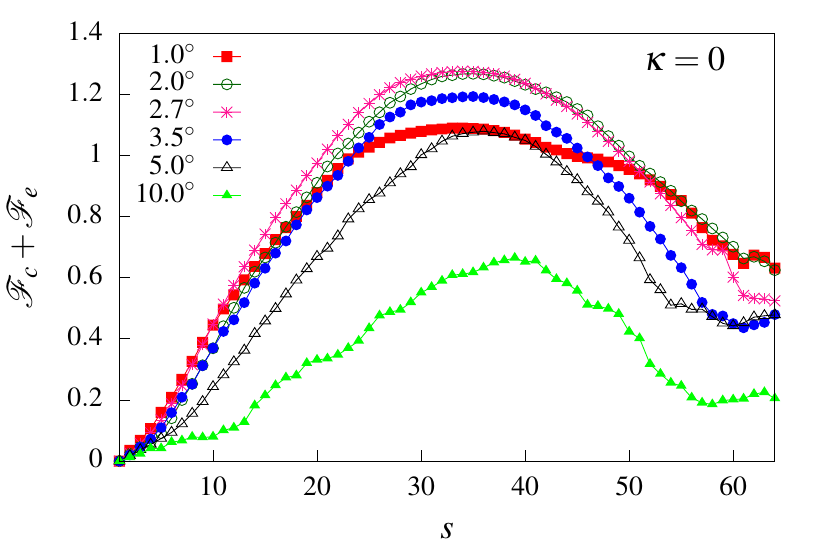}}\hspace{\fill}\quad 
\subfloat[]{\includegraphics[width=0.48\textwidth]{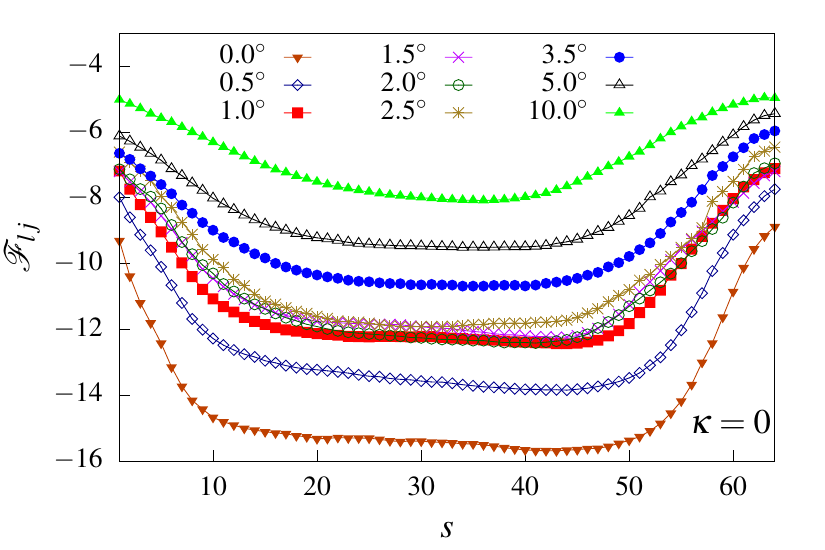}}\\
\subfloat[]{\includegraphics[width=0.48\textwidth]{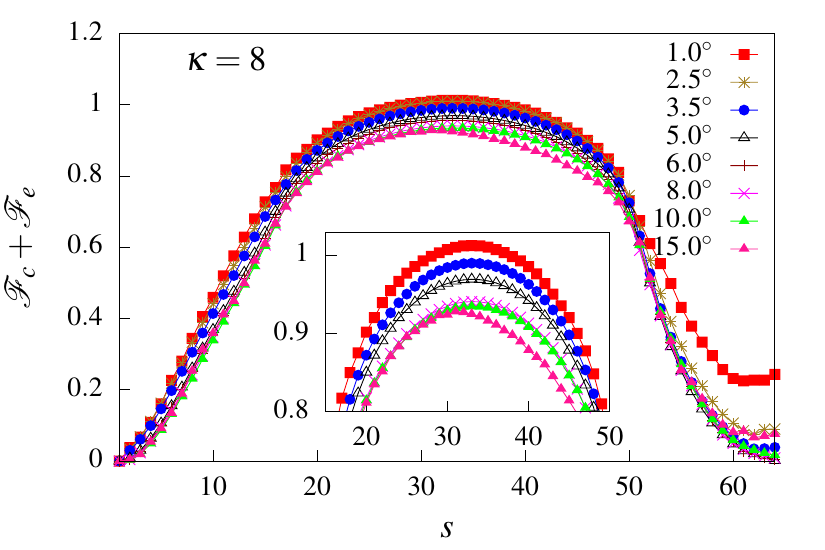}}\hspace{\fill}\quad 
\subfloat[]{\includegraphics[width=0.48\textwidth]{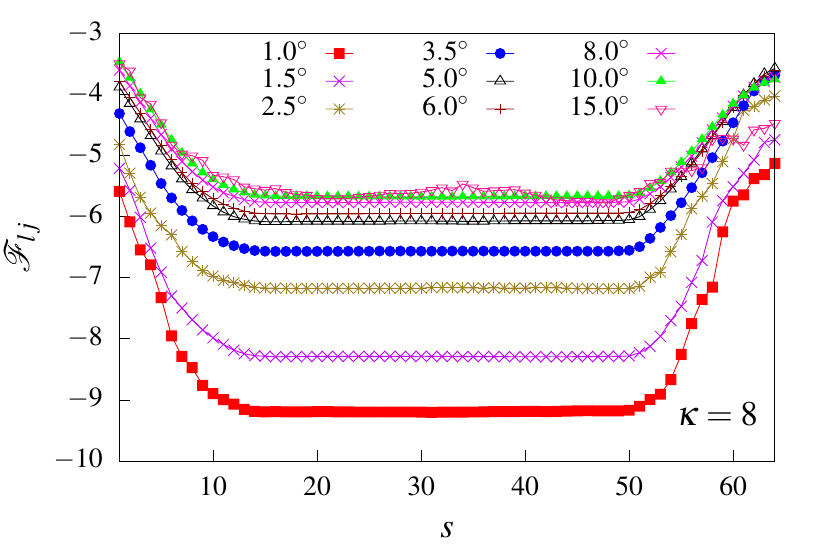}}
\caption{(a) Entropic part of the free energy ($\mathscr{F}_c + \mathscr{F}_e$) for a flexible polymer as a function of the translocation coordinate $s$ for different $\alpha$ values showing non-monotonic behavior. (b) Energetic contribution to free energy for a flexible polymer from surface interactions ($\mathscr{F}_{lj}$) as a function of translocation coordinate $s$ for various $\alpha$ values.  (c) Entropic part of the free energy ($\mathscr{F}_c + \mathscr{F}_e$) for a semiflexible polymer ($\kappa = 8$) as a function of the translocation coordinate $s$ for different $\alpha$ values showing largely monotonic behavior. (Inset) The barrier height decreases monotonically with $\alpha$. (d) Energetic contribution to free energy for the semiflexible polymer from surface interactions ($\mathscr{F}_{lj}$) as a function of translocation coordinate $s$ for various $\alpha$ values.
}\label{freeE0}
\end{figure*}

It has been established that the relaxation time of a polyelectrolyte chain in equilibrium is much shorter than the translocation time for the typical range of chain lengths used and typical voltages applied in most experiments~\cite{katkar2018role}. This implies that a polymer chain initially at equlibrium can be taken to be in quasi-equilibrium as it translocates across the nanochannel especially for low external forces. In fact the quasi-equlibrium approximation in such cases can be even extended to translocation of every segment of the polymer~\cite{katkar2018role}. The quasi-equlibrium approximation obviously fails for high forces and longer lengths of the polymer. Using this approximation for low forces and chain lengths considered in our simulations, we construct the free energy landscape for the translocation process to understand the translocation behavior obtained in our simulations.

There are several contributions to the free energy :  (i) the entropic contribution coming from the confinement of the polymer segment inside the channel (ii) entropic contribution due to the polymer segments outside the channel (ii) energetic contribution due to surface interactions between the channel and the polymer and (iv) energetic contribution due to the external force. 
The total free energy of the system in terms of the translocation coordinate can be written as
\begin{equation}
     \mathscr{F}_{t}(s) =  \mathscr{F}_{e}+ \mathscr{F}_{c}+ \mathscr{F}_{pp}
\end{equation}
where $ \mathscr{F}_c$ is the confinement free energy, $ \mathscr{F}_e$ is the entropic component due to polymer segments outside the channel and $ \mathscr{F}_{pp}$ is the free energy contribution due to the channel polymer interactions.

To construct $ \mathscr{F}_e$ for a semiflexible polymer segment inside a conical channel, we use the approach in~\cite{nikoofard2013directed}. The contribution to the free-energy from a polymer of size $R$ confined in an area $A$ of the cone in two-dimensions is given as $\mathscr{F}_c \sim k_BT \left(\frac{R^2}{A}\right)^{\frac{1}{2\nu - 1}}$  \cite{sakaue2006polymer}. If $N_p$ denotes the number of polymer beads confined inside the channel, then $R \sim \sigma N_p^\nu$, where $\nu = 3/4$ is the Flory exponent in two-dimensions. $A$ is dependent on the degree of confinement. If $L$ denotes the length of the region inside the channel along the $x-$direction that holds the confined polymer and $a$ the length which is empty, then $A$ is given as $A = \frac{1}{4\tan\alpha}\left[(D_0+2(a+L)\tan \alpha)^{2}-(D_0+2a\tan \alpha)^{2}\right]$.
 %

In the case of rigid polymers, the analysis can be extended by replacing the number of monomers $N_p$ with the number of Kuhn segments $\tilde{N_p} \approx N_p\sigma/l_{K}$ where $l_K = 2l_p$ is the Kuhn length. $l_p$ is the persistence length of the polymer. In two-dimensions, the bending rigidity is related to the persistence length as $\kappa/k_BT = l_p/2$. We would however like to express $\mathscr{F}_e$ in terms of the translocation coordinate $s$. Therefore, we extract the number of beads that are inside the channel from the simulations for a fixed translocation coordinate. As is expected, there would be several configurations of the polymer (with different values of $N_p$) inside the channel for a given $s$. The confinement free energies of all such configurations are averaged to give $ \mathscr{F}_c(s)$. 





The free energy contribution from the polymer segments outside the channel is given as\cite{gauthier2008monte2}: 
\begin{equation*}
        \mathscr{F}_{e} = \begin{cases}
                        \gamma^{\prime}\log N_{c} & \text{ $N_{c}>1$ and $N_{t}=0$}\\
        \gamma^{\prime}\log N_{c} + \gamma^{\prime}\log N_{t} & \text{ $N_{c}>1$ and $N_{t}>1$}\\
                            \gamma^{\prime}\log N_{t} &\text{ $N_{c}=0$ and $N_{t}>1$}
                        \end{cases}
\end{equation*}
where $N_{c}$ is the number of beads on the cis-side of the channel and $N_{T}$ is the number of beads on the trans-side of the channel. Note that $N_c$, $N_T$ are functions of the translocation coordinate $s$ and are obtained from the simulation similar to the confinement free energy for every value of $s$. Further, $N_c + N_p + N_T = N$ at all times. 

The free energy contribution from the surface interaction $\mathscr{F}_{lj}$ is obtained by summing over the LJ potential felt by each polymer bead when they are inside the channel. For a given translocation coordinate, we identify the number of beads inside the channel and calculate the total potential felt by the beads. As we mentioned before, for a given $s$, there would be multiple configurations of the polymer inside the channel (corresponding to the time steps for which $s = 1, 2, ...$). All such contributions are appropriately accounted for and the averaged free energy contribution $\mathscr{F}_{lj}$ due to surface interactions is evaluated numerically as the translocation coordinate changes. 

In Fig.~\ref{freeE0}(a), we plot the total free energy contributions which are entropic in origin for a flexible polymer, arising due to the confined segments of the polymer and the segments outside the channel. The free energy contribution due to channel-polymer interactions are plotted in Fig.~\ref{freeE0}(b). The entropic contribution exhibits a free energy barrier for an intermediate $s$ for all values of the channel angle $\alpha$. This indicates that entropically it is more difficult for the polymer to enter the channel from the cis side. However, the energetic contribution coming from the attractive $LJ$ potential dominates the initial stages and the polymer is sucked into the channel. At intermediate $s$, the competition between the two leads to the comparatively larger waiting times observed in Fig.~\ref{fig:forwardresk0}. As $\alpha$ is increased, we observe a clear non-monotonic behavior in the entropic part of the free energy. The free energy barrier increases as we go from $\alpha = 1.0^{\circ}$ to $\alpha = 2.7^{\circ}$ and decreases as $\alpha$ is increased further. This indicates that translocation would be more difficult from $\alpha = 1.0^{\circ}$ to $\alpha = 2.7^{\circ}$ and favorable as the channel angle increases. This is consistent with the behavior observed in the translocation time as a function of the channel angle, where for $f_0 = 0.1$, $\tau$ shows a peak at $\alpha \approx 2.7^{\circ}$ and decreases with increasing $\alpha$. The energetic contribution $ \mathscr{F}_{lj}$ does not show non-monotonic behavior at these values of $\alpha$, with significant overlap in the free energy profiles (see Fig.~\ref{freeE0}(b)). As $\alpha$ is increased, $ \mathscr{F}_{lj}$ becomes shallower indicating easier translocation and therefore decreasing $\tau$.   

In Fig.~\ref{freeE0}(c), we plot the total free energy contributions which are entropic in origin for a semiflexible polymer for various channel angles. Similar to the flexible polymer, the entropic contribution indicates a barrier at intermediate $s$. However, unlike the flexible polymer, this barrier reduces monotonically with increasing $\alpha$. The corresponding free energy contribution due to channel-polymer interactions also becomes shallower monotonically as $\alpha$ is increased (see Fig.~\ref{freeE0}(d)). Therefore, for a semiflexible polymer, translocation becomes easier with increasing $\alpha$. If we compare our analytical predictions with that of the simulation results in Fig.~\ref{rigidtrans}(a) for $\kappa = 8$, we find that translocation time indeed decreases with increasing $\alpha$ upto $\alpha = 6^{\circ}$ consistent with our predictions. However, for larger channel angles, the translocation time increases. This behavior is very different from the predictions from the free energy calculations. 

In order to understand this behavior, it is crucial to look at the potential landscape inside the channel. At large channel angles, the strength of the potential due to the channel walls becomes negligible for large regions inside the channel. Even during entry of the polymer from the cis side of the channel, the attractive potential becomes much weaker in the region close to the cis side for large angles. This hampers the pulling ability of the channel resulting in larger waiting times of even the initial beads of the polymer. As the polymer is pulled in, the segments of the stiff polymer inside the channel keeps getting deflected from one wall to the other (see Supplementary Movie). The potential near the centre of the channel is negligible as is the external force and the polymer segments stay attached to either wall for considerable times. This leads to increasing translocation times at higher channel angles. For smaller channel angles, the average positions of the polymer segments are much closer to the middle of the channel as both walls try to pull on them. With external force much stronger near the center, the polymer is pulled out faster. Note that for flexible polymers, the possibility of more conformations of the polymer as compared to stiffer chains at larger angles ensures that the situation is much more homogeneous and we do not observe much change at larger channel angles.

\section{Discussions}
The translocation of polymers through asymmetric shaped channels has sparked a lot of interest due to its significant advantages in bio-sensing applications. Buoyed by experimental studies using conical nanochannels, we have studied the translocation dynamics of a semiflexible polymer through such a channel under the influence of a spatially varying external drive and attractive surface interactions. The waiting time distribution show rich features arising due to the polymer stiffness, surface interactions and nature of the external drive. We attempt to understand some of these features using a free energy argument based on a quasi-equilibrium approximation which is applicable for smaller polymer lengths and low forces. The theory captures some of the non-monotonic features of the translocation dynamics. At higher stiffness and larger channel angles, the variable potential landscape arising from the surface interactions becomes dominant. This variability needs to be accounted for in order to provide a more accurate description of the dynamics.  

In our analysis we have ignored the role of electrostatic interactions~\cite{buyukdagli2019theoretical}. Translocation rate depends on the charge density of the polymer and the density of the bulk concentration of the solvent in a confined pore. Low polymer charge density and low solvent concentration leads to lower translocation rates, while translocation rate increases sharply for high polymer charge density at high solvent concentration. It will be interesting to ask how these rates could be modified when considering asymmetric channels like the one considered in this study.


Recent experiments have revealed that driven polymer translocation through synthetic nanopores is a two stage process with translocation initially slowing with time before accelerating close to the end of the process~\cite{chen2021dynamics}. In our detailed simulations, we show however that the translocation dynamics is strongly dependent on the stiffness of the polymer and surface interactions. Attractive surface interactions can considerably slow down the translocation process near the ends while increasing the angle of the conical pore may facilitate the process.  We have considered one possible conical structure where the polymer enters the pore from the constricted side of the channel. In future work we will explore the detailed dynamics of the polymer as it enters from the wider end of the channel which is expected to give rise to significantly different dynamics due to the asymmetry of the channel. 


\section{ACKNOWLEDGEMENTS:}
AS acknowledges  University Grants Commission (UGC), India for the financial support under the UGC-JRF/SRF scheme. We are thankful to the computational facility provided by IISER, Mohali. 

\bibliography{references}

\end{document}